\documentstyle[aps]{revtex}
\input epsf

\begin{document}
\title{Velocity correlations and the structure of nonequilibrium hard core fluids}
\author{James F. Lutsko}
\address{Physics Division, Starlab \\
Rue Engelandstraat 555 \\
B-1180 Brussels, Belgium \\
and \\
Center for Nonlinear Phenomena and Complex Systems\\
Universit\'{e} Libre de Bruxelles\\
1050-Bruxelles, Belgium
email: jim.lutsko@skynet.be}
\date{\today}
\maketitle

\begin{abstract}
A model for the pair distribution function of nonequilibrium hard-core
fluids is proposed based on a model for the effect of velocity correlations
on the structure. Good agreement is found with molecular dynamics
simulations of granular fluids and of sheared elastic hard spheres. It is
argued that the incorporation of velocity correlations are crucial to
correctly modeling atomic scale structure in nonequilibrium fluids.
\end{abstract}


The structure of a fluid as characterized by the pair distribution function
(pdf)\ plays a central role in statistical mechanics. In equilibrium, it
allows one to calculate the equation of state whereas away from equilibrium
it is required, in the guise of the equal time density-density correlation
function, whenever one wishes to project a kinetic equation onto the
hydrodynamic subspace, see e.g. ref\cite{Lutsko_ShearFluctuations}. In both
cases, it can be directly measured by light scattering\cite{Reichl} and is
thus an interesting quantity in order to make connections with experimental
studies. The theory of the structure of equilibrium fluids is quite advanced
and ranges from simple models for spherical hard core systems to
perturbative models applicable to continuous potentials to integral
equations applicable to a wide range of systems\cite{HansenMcdonald},\cite
{Rogers84}. Much less is known for nonequilibrium fluids. Kinetic theory can
yield information in certain regimes such as low density or large spatial
separations, see. e.g. \cite{Mirim},\cite{Lutsko_Fluctuations} but does not
provide complete models of nonequilibrium structure applicable to dense
fluids. Phenomenological models exist 
 based on fluctuating hydrodynamics\cite
{RonisShear},\cite{Ernst1}\cite{Ernst2} or Langevin models\cite{Rainwater83},\cite{Hess85} but
these can only be expected to be applicable at large length scales.
For moderately dense hard-core fluids, the only realistic and tractable
kinetic theory is the Enskog equation describing the time-evolution of the
one-body distribution function (the dense fluid generalization of the
Boltzman equation)\cite{Beijeren79}. Recently, it has been shown that one
piece of structural information is directly accessible in hard-core systems
within the same set of approximations used to derive the Enskog kinetic
theory - namely, the pair distribution function at contact\cite{Lutsko96}.
The purpose of this Letter is to show that it is possible to use this
information to extend the simplest class of equilibrium models, the
so-called Generalized Mean Spherical Approximation or GMSA\cite{WaismanGMSA}%
, \cite{HansenMcdonald}, to nonequilibrium systems and to validate the
models constructed against molecular dynamics simulations. In particular,
the models will be compared in the case of a granular fluid ( a homogeneous
system violating energy conservation) as well as strongly sheared elastic
hard spheres. The latter case is particularly interesting due to the fact
that it has motivated previous attempts to model nonequilibrium structure,
is a realistic model of some sheared colloidal suspensions and has been
investigated experimentally for such systems\cite{Ackerson88}.

To construct a model for the structure of a {\it nonequilibrium} fluid, we begin, 
as in nearly all models of equilibrium structure, with the Ornstein-Zernike (OZ) equation for a uniform fluid
\begin{equation}
h({\bf r}_{1},{\bf r}_{2})=c({\bf r}_{1},{\bf r}_{2})+\rho \int d{\bf r}%
_{3}\;c({\bf r}_{1},{\bf r}_{3})h({\bf r}_{3},{\bf r}_{2})  \label{OZ}
\end{equation}
where $\rho $ is the density, the structure function $h({\bf r}_{1},{\bf r}%
_{2})$ is related to the pdf, $g({\bf r}_{1},{\bf r}_{2})$, by $h(%
{\bf r}_{1},{\bf r}_{2})=g({\bf r}_{1},{\bf r}_{2})-1$ and eq.(\ref{OZ})
serves as a definition of the direct correlation function (dcf)$\;c({\bf r}%
_{1},{\bf r}_{2})$. By definition, hard core atoms cannot interpenetrate so
the probability of finding two atoms closer together than the hard sphere
diameter, $\sigma $, is zero giving the boundary condition $\Theta \left(
\sigma -r_{12}\right) g({\bf r}_{1},{\bf r}_{2})=0$ where $r_{12}=\left| 
{\bf r}_{1}-{\bf r}_{2}\right| $ and $\Theta (x)$ is the Heaviside step
function which is equal to 1 for $x>0$ and zero otherwise. Various
arguments, based e.g. on the Mayer expansion of the various quantities,
leads to the conclusion that the direct correlation function is short ranged
and the Percus-Yevick (PY)\ theory results from taking it to be zero outside
the core:\ $\Theta \left( r_{12}-\sigma \right) c({\bf r}_{1},{\bf r}_{2})=0$%
. This is sufficient to completely specify all quantities and, in
particular, the pdf can be analytically determined\cite{HansenMcdonald}.
However, there is ambiguity in the equation of state since it may be
determined from either the pressure equation, which involves the pdf at
contact, or from the compressibility equation involving an integral of the
structure function over all space: these do not yield the same result. A
feature of many, if not most, successful theories of equilibrium structure
is that they force both routes to the equation of state to yield the same
result, a requirement known as thermodynamic consistency. The GMSA
accomplishes this by giving the direct correlation function a parameterized
''tail'' by requiring $\Theta \left( r_{12}-\sigma \right) c({\bf r}_{1},%
{\bf r}_{2})=v({\bf r}_{1},{\bf r}_{2})$ for some function $v({\bf r}_{1},%
{\bf r}_{2})$. In the original MSA, the tail is taken to be either the
interatomic potential or the Meyer function constructed from it (thus
reducing to the PY approximation in the case of hard spheres). In the
original GMSA of Waisman\cite{WaismanGMSA}, the tail is taken to be a Yukawa 
$v({\bf r}_{1},{\bf r}_{2})=Ae^{-B(r_{12}-1)}/r_{12}$ and the two parameters
are determined by requiring that both the routes to the equation of state
yield the Carnahan-Starling equation. The pressure equation therefore fixes
the pdf at contact whereas the compressibility equation fixes its area. This
model may be solved analytically and gives a good model of the pdf even for
dense fluids and can be adapted to give reasonable models other systems such
as plasmas\cite{PlamerWeeks73}. In all cases, the two elements which depend
on the system being in equilibrium are (1)\ the arguments for the nature of
the direct correlation function outside the core and (2)\ the equations of
state and the condition of thermodynamic consistency.

The first result that suggests the applicability of these models away from
equilibrium is the work of Yuste and Santos on the structure of more complex
hard-core systems\cite{Yuste91}. They show
that the pdf can be written in terms of an auxiliary function which is then
parameterized as a Pad\'{e} approximant. This approximant is then taken to
be the simplest possible that satisfies three physical conditions:\ the pdf
is finite at contact; the pdf goes to one at large separations; the Fourier
transform of the structure function is finite at all wavevectors. These
conditions are sufficient to completely specify the auxiliary function and
the result is the PY\ pdf. Extending the approximant by one additional term
in the numerator and denominator (the relative order of the two is fixed by
the constraints) is exactly equivalent to the Yukawa closure and the
imposition of thermodynamic consistency and the Carnahan-Starling equation
of state yields the GMSA. By recasting the GMSA solely in terms of general
physical requirements and extendable analytic approximations, this work
accentuates the fact that insofar as the tail of the direct correlation
function is simply being parameterized, the GMSA\ is generic. However,
before this model can be used in nonequilibrium systems, some substitute for
the equation of state, used as input, must be found. One role of the
equation of state is, as mentioned above, to fix the value of the pdf at
contact. The second ingredient needed to complete the model is the recent
result showing that it is possible to calculation any two-body function,
evaluated for two atoms in contact, with the same degree of approximation as
is used in the Enskog equation (the dense fluid generalization of the
Boltzmann equation for hard core systems)\cite{Lutsko96}. A slightly more
general form of the result than that given in ref\cite{Lutsko96} is that the
(nonequilibrium)\ average value of any two-body function at contact, say $U(%
{\bf r})=\left\langle \sum_{i<j}u(ij)\delta \left( {\bf q}_{ij}-{\bf r}%
\right) \right\rangle $ where we use the abbreviated notation $u(ij)\equiv u(%
{\bf q}_{i},{\bf q}_{j},{\bf p}_{i},{\bf p}_{j})$, is given by 
\begin{eqnarray}
U(\sigma \widehat{{\bf r}}) &=&\left\langle \sum_{i<j}\delta \left( {\bf q}%
_{ij}-\sigma \widehat{{\bf r}}\right) u(ij)\right\rangle _{0}  \nonumber \\
&&+\left\langle \sum_{i<j}\delta \left( {\bf q}_{ij}-\sigma \widehat{{\bf r}}%
\right) 
\left( \Theta \left( -\widehat{{\bf q}}_{ij}\cdot {\bf p}_{ij}\right)\frac{1}{\alpha }u^{\prime }(ij)-\Theta \left( \widehat{{\bf q}}_{ij}\cdot {\bf p}_{ij}\right)u(ij)\right) \right\rangle _{0}
\label{contact}
\end{eqnarray}
where the parameter $\alpha $ characterizes the inelasticity of the hard
spheres ($\alpha =1$ corresponds to elastic hard spheres), $\widehat{{\bf r}}%
={\bf r}/r$ denotes a unit vector and  $u^{\prime }({\bf q}_{i},{\bf q}_{j},%
{\bf p}_{i},{\bf p}_{j})=u({\bf q}_{i},{\bf q}_{j},{\bf p}_{i}-\frac{%
1+\alpha }{2}\widehat{{\bf q}}_{ij}\left( \widehat{{\bf q}}_{ij}\cdot {\bf p}%
_{ij}\right) ,{\bf p}_{j}+\frac{1+\alpha }{2}\widehat{{\bf q}}_{ij}\left( 
\widehat{{\bf q}}_{ij}\cdot {\bf p}_{ij}\right) )$ is the value of the
two-body function just after the collision has ocured. (This result is
easily obtained by applying the method of ref.\cite{Lutsko96} to the
pseudo-Liouville equation for inelastic hard spheres\cite{GranularLiouville}%
). The notation $\left\langle ...\right\rangle _{0}$ indicates an average
that it is to be performed with the factorized two-body distribution
function $f_{2}({\bf q}_{i},{\bf q}_{j},{\bf p}_{i},{\bf p}_{j})=f_{1}({\bf q%
}_{i},{\bf p}_{i})f_{1}({\bf q}_{j},{\bf p}_{j})g_{0}({\bf q}_{i},{\bf q}%
_{j})$ where $f_{1}({\bf q}_{i},{\bf p}_{i})$ is the one-body distribution
(normally derived from the Enskog equation) and $g_{0}({\bf q}_{i},{\bf q}%
_{j})$ is typically the local-equilibrium pdf. The first term on the right 
represents the average over the nonequilibrium state when all velocity
correlations are neglected whereas the second gives the effect of velocity
correlations generated by the collision. As discussed previously\cite
{Lutsko96}, the Enskog equation also takes account of these velocity
correlations whereas both it and eq.(\ref{contact}) neglect correlations
that persist for more than one collision.

In summary, I therefore propose to use a GMSA-like paramterized closure of the OZ equation and
to fix the parameters using information coming from equation (2) which, in the case 
$u(ij)=1$ gives a generalization of the pressure equation applicable to nonequilibrium fluids.
The form of the parameterization of the tail of the dcf will be guided from results from
kinetic theory: the Yukawa form will be used when the pdf is expected to decay exponentially,
as in the case of a granular system \cite{Ernst2} (note that this corrects a previously
reported algebraic decay \cite{Ernst1} for this system) whereas a power-law will be used
when the decay is expected to be algebraic (as for shear flow).

\begin{figure}[t]
\begin{center}
\leavevmode
\epsfxsize=5in 
\epsfbox{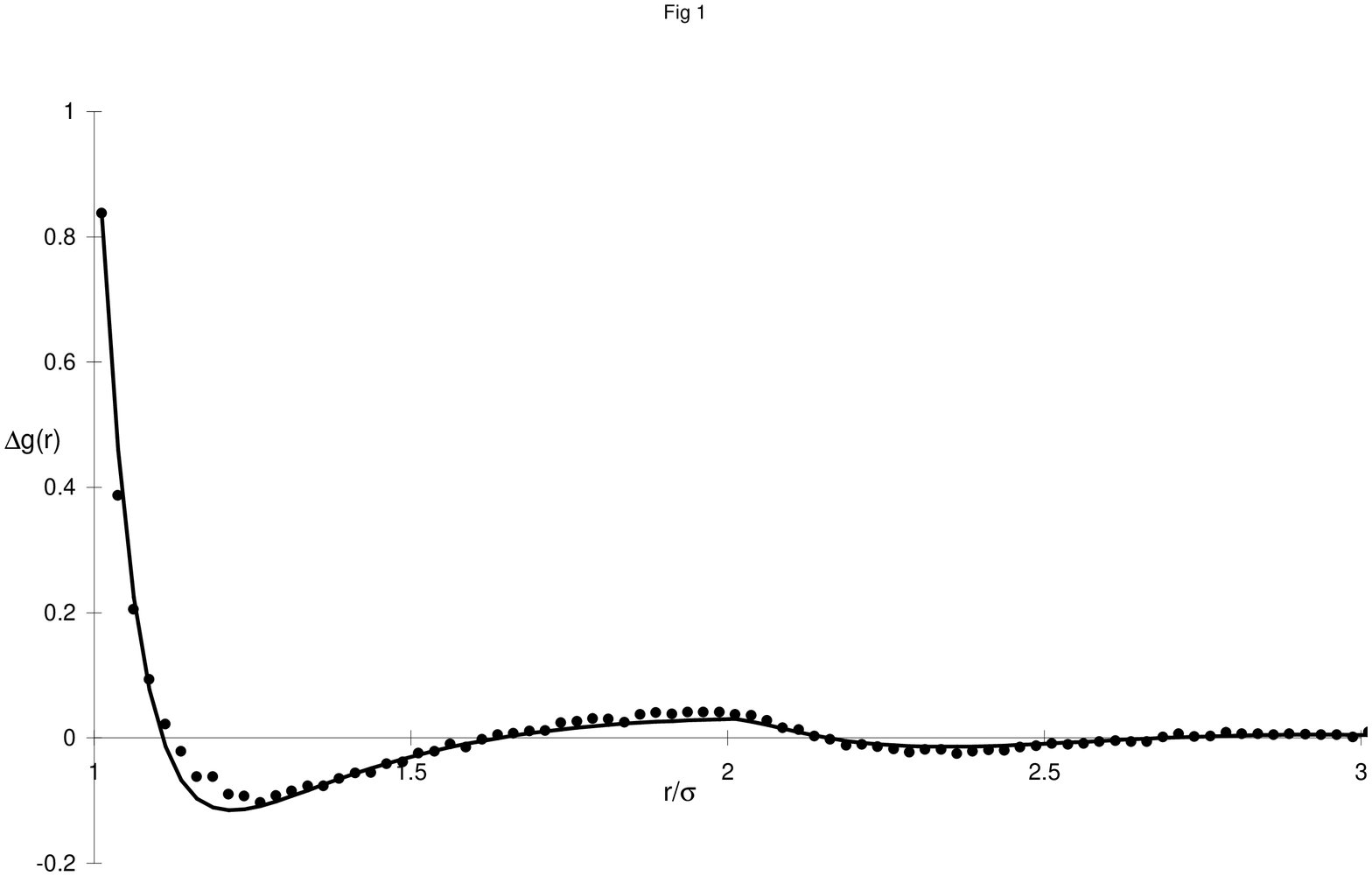}
\end{center}
\caption{The difference between the pdf for a model granular fluid with $n\sigma^3=0.5$
and $\alpha=0.5$ and the PY pdf for an equilibrium fluid of the same density. The circles are
from simulation and the line is the GMSA result.}
\label{fig1}
\end{figure}

Granular fluids are often modelled as hard spheres which lose energy upon collision in which case 
the undriven system cools at a steady rate. The energy loss per collision, and hence the cooling rate,
is controlled by the inelasticity parameter, $\alpha$, introduced in eq.(\ref{contact}) above. 
In this homogeneous cooling state (HCS)\ eq.(\ref{contact})\ gives $%
g(r_{12}=\sigma ;\alpha )=\frac{1+\alpha }{2\alpha }\chi $ where $\chi $ is
the equilibrium pdf at contact (since the HCS is translationally invariant,
all two-body quantities depend only on the relative separation). The
validity of eq.(\ref{contact}) for this system has been studied in some
detail by means of a set of molecular dynamics simulations which will be
presented at a later time.  The simulations show that it is a reasonable
approximation in the range $1>\alpha \succsim 0.5$ at which point
mode-coupling effects become important. Here, we only present one
comparison, figure(\ref{fig1}), between the pdf as determined from
simulation and the nonequilibrium GMSA with the Yukawa closure for a
relatively dense fluid, $n\sigma ^{3}=0.5$, and a large rate of dissipation, 
$\alpha =0.5$. The model is seen to be good not only near the core but also
in its representation of the decay of the nonequilibrium contribution to the
structure.  A measure of the significance of the nonequilibrium effects shown is that they
are comparable in magnitude to the equilibrium pdf which, for this
density, varies between a maximum of 2.16 (at the core) and zero, with an asymptotic value of 1.
In applying the GMSA in this case, we use the known analytic
solution of this model\cite{Hoye77}, and must determine the two parameters
introduced by the Yukawa tail by the constraints. One constraint comes from
the value of the pdf at contact but there is no obvious replacement for the
compressibility equation used in equilibrium so we continue to enforce it,
using the expression for the pressure given in ref.\cite{Garzo99}. Some
experimentation shows that the model is relatively insensitive to the latter
approximation. A more realistic constraint, to be pursued in a future publication, would
be to replace the compressibility equation by the requirement that the area under the 
pdf agree with that obtained from fluctuating hydrdodynamics calculations\cite{Ernst2}.

A second, and more complex test, is the application of the nonequilibrium
GMSA to shear flow. Define a cartesian coordinate system in which the
macroscopic flow field is ${\bf v}({\bf r})=ay\widehat{{\bf x}}$, where $a$
is the shear rate. In this case, an expression for the pdf at contact has
been studied in some detail\cite{Lutsko96} and, while it is possible to
calculate its full angular dependence, in the illustrative calculation
presented here, we will use the simple approximation $g(\sigma {\bf r}%
;a)=\chi \left( 1+A\widehat{{\bf r}}_{x}\widehat{{\bf r}}_{y}\right) $ with $%
A$ chosen to give the calculated value at $\widehat{{\bf r}}_{x}=\widehat{%
{\bf r}}_{y}=\sigma /\sqrt{2}$\cite{Lutsko96}, which is the dominant
contribution. To model this, we must use an anisotropic tail for the dcf and
therefore take $v({\bf r}_{1},{\bf r}_{2})=v_{0}(r_{12})+v_{1}(r_{12})%
\widehat{{\bf r}}_{12x}\widehat{{\bf r}}_{12y}$. \ The solution of the OZ equation 
with an anisotropic potential is well known \cite{Gray}, \cite{HansenMcdonald} and will
only be sketched here. To begin, the structure function and the dcf must be
expanded in terms of their angular dependencies:\ $h({\bf r}%
_{12})=h_{lm}(r_{12})Y_{lm}(\widehat{{\bf r}}_{12})$ where repeated indices
are summed and $Y_{lm}(\widehat{{\bf r}})$ is a spherical harmonic. We will
also need the Fourier transforms of these expansions which takes the form $%
\widetilde{h}({\bf k})=\overline{h}_{lm}(k)Y_{lm}(\widehat{{\bf k}})$ where $%
\overline{h}_{lm}(k)$ may be calculated from $h_{lm}(r_{12})$ via a Hankel
transformation (in the case $l=0$ this reduces to a Fourier transform).
Fourier transforming the OZ equation and multiplying by $Y_{ij}^{\ast }(%
\widehat{{\bf k}})$ and integrating then gives an infinite set of coupled
equations. The minimal set of components that we can work with are the
spherically symmetric component, $h_{00}(r_{12})$, and the components that
combine to produce the angular dependence of the tail,  $\widehat{{\bf r}}%
_{12x}\widehat{{\bf r}}_{12y}$, namely $h_{22}(r_{12})$ and $h_{2-2}(r_{12})$%
. Furthermore, we know that at contact, these give the dominant
contributions to $h({\bf r}_{12})$. For illustrative purposes, we therefore
keep only these three components and the corresponding equations in the
hierarchy with the result that we must solve the system 
\begin{eqnarray}
\overline{h}_{00} &=&\overline{c}_{00}+n\sqrt{\frac{1}{4\pi }}\left[ 
\overline{c}_{00}\overline{h}_{00}+\overline{c}_{22}\overline{h}_{2-2}+%
\overline{c}_{2-2}\overline{h}_{22}\right]   \nonumber \\
\overline{h}_{22} &=&\overline{c}_{22}+n\sqrt{\frac{1}{4\pi }}\left( 
\overline{c}_{00}\overline{h}_{22}+\overline{c}_{22}\overline{h}_{00}\right) 
\nonumber \\
\overline{h}_{2-2} &=&\overline{c}_{2-2}+n\sqrt{\frac{1}{4\pi }}\left( 
\overline{c}_{00}\overline{h}_{2-2}+\overline{c}_{2-2}\overline{h}%
_{00}\right) 
\end{eqnarray}
together with the boundary conditions described above. Furthermore, a
consideration of the symmetry of the problem and the form of the boundary
conditions shows that $\overline{h}_{22}$ must be pure imaginary and that $%
\overline{h}_{22}=\overline{h}_{2-2}^{\ast }$ so that the independent
equations can be written as 
\begin{equation}
\left( \overline{h}_{00}\pm i\overline{h}_{22}\right) =\left( \overline{c}%
_{00}\pm i\overline{c}_{22}\right) +n\sqrt{\frac{1}{4\pi }}\left( \overline{h%
}_{00}\pm i\overline{h}_{22}\right) \left( \overline{c}_{00}\pm i\overline{c}%
_{22}\right)   \label{factored}
\end{equation}
which has the form of two decoupled OZ equations (at this point, the solution maps
to that of the dipolar hard-sphere model \cite{HansenMcdonald}). The full pdf now has the
form $g({\bf r})=1+\frac{1}{\sqrt{4\pi }}h_{00}(r)+i\sqrt{\frac{15}{2\pi }}%
h_{22}(r)\widehat{{\bf r}}_{x}\widehat{{\bf r}}_{y}$. The only element that remains is to specify the tails of the
dcf. At contact, we know that $h_{00}(r)$ should be unchanged from its
equilibrium value whereas $h_{22}(r)$ is substantial. This suggests that we
take $v_{0}(r_{12})=0$ although in a more sophisticated treatment we might
want to use a Yukawa so as to fix the value of $h_{00}(r)$ at contact (from
the model for the pdf at conact, this should be unchanged from equilibrium
but eq.(\ref{factored}) will not necessarily respect this - however, upon solving the model, 
the change of this component at contact is found to be minimal). For the
anisotropic part, the simplest choice we can make is to take $%
v_{1}(r_{12})=B/r_{12}^{3}$ (since then these equations may be solved using
the PY results\cite{HansenMcdonald}). Although kinetic theory calculations
\cite{Mirim} indicate the presence of a long-ranged tail for $h_{22}(r)$
that decays as $1/r$, we nevertheless proceed using $%
v_{1}(r_{12})=B/r_{12}^{3}$ because (a) even using a longer-ranged $1/r$
tail for the dcf will not yield $1/r$ behavior in the pdf\cite{PlamerWeeks73}
and (b) no matter what tail we choose, there will be a $1/r^{3}$ decay\cite
{HansenMcdonald}. The constant $B$ is determined so as to give agreement with 
the predicted value of the pdf at contact and since there are no other parameters, 
the ad hoc use of the compressibility equation is not needed in this case. The solution then
follows closely that for dipolar hard spheres, see e.g. \cite{HansenMcdonald}%
. To test this, I have performed molecular dynamics computer simulations,
following the procedures described in ref.\cite{Lutsko96},  to determine the
functions $h_{00}(r)$ and $h_{22}(r)$ and a comparison to the nonequilibrium
GMSA model is shown in Figs. (2) and (3) for a moderately dense fluid, $%
n\sigma ^{3}=0.5$, in a strongly driven state, $at_{00}=0.5$ where $%
t_{00}=\sigma /\left( 4n^{\ast }\sqrt{\pi k_{B}T}\right) $ is the Boltzmann
collision time. For $h_{00}(r)$, there is little difference between the
theoretical result, which is virtually identical to the PY pdf, and the
simulation indicating little or no change to this projection of the pdf due
to the shear flow. For the more interesting $h_{22}(r)$ component, the
agreement is not exact but this crude calculation nevertheless gives a
reasonable approximation to the amplitude and peak of the oscillations of
this quantity which arises solely from the nonequilibrium state and which are seen by comparison
to fig. (2) to be comparable in magnitude to the equilibrium pdf. A more
complete solution to this model is in preparation.

These results on two very different strongly nonequilibrium systems show
that the knowledge of the velocity correlations produced when two atoms
collide, and the consequent contribution to the pdf, is sufficient to
generalize a simple class of structural models to nonequilibrium fluids.
While these models may be useful as input to kinetic theory or for the
interpretation of light-scattering experiments, the most important point to
be made here is the critical role played by the velocity correlations in
determining the atomic-scale structure in nonequilibrium systems. Theories
of nonequilibrium structure which do not explicitly take into account
atomic-scale velocity correlations, such as the phenomenological theories of
Hess\cite{Rainwater83},\cite{Hess85} and Ronis\cite{RonisShear} and the
semi-phenomenological theory of Eu\cite{EuPDF}, are, at least when applied
to hard-core systems, {\it a priori} unlikely to be able to model the
small-length scale distortions of the fluid structure produced by
nonequilibrium processes while, as shown above, even the simplest theories
which do take them into account compare reasonably well to simulation.

\begin{figure}[t]
\begin{center}
\leavevmode
\epsfxsize=5in 
\epsfbox{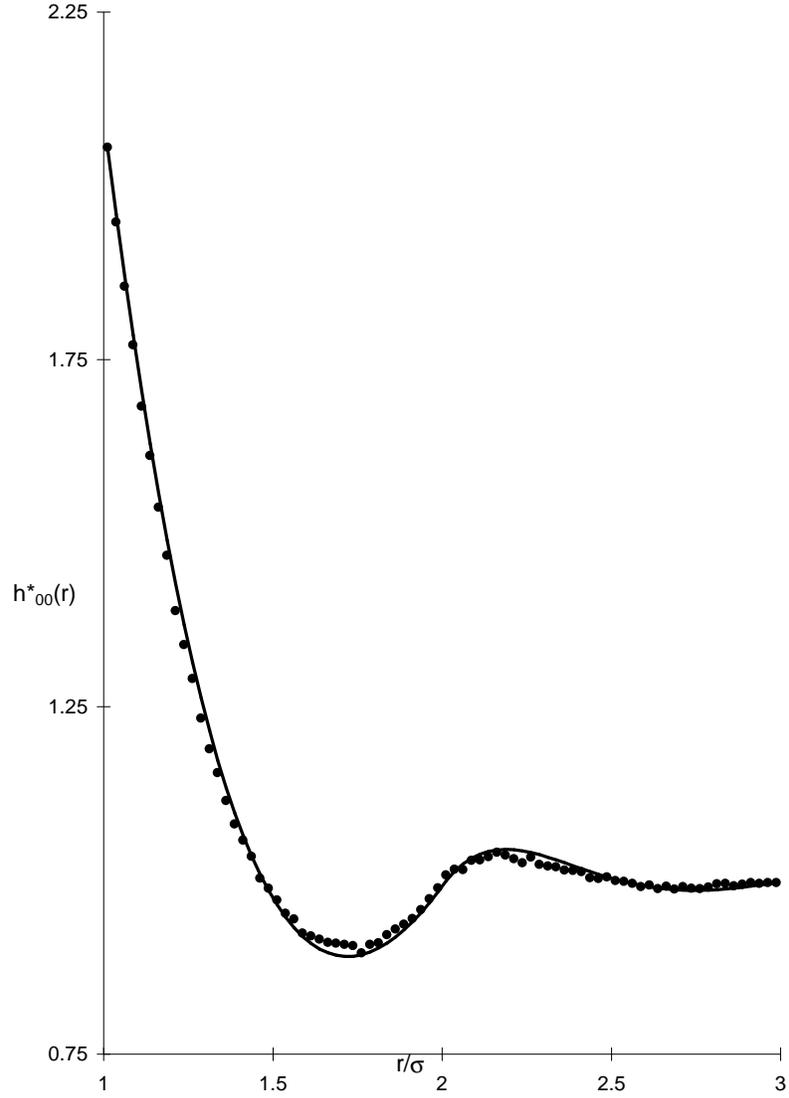}
\end{center}
\caption{The quantity $h^{*}_{00}(r)=\frac{1}{\sqrt{4\pi }}h_{00}(r)$ for a fluid of sheared hard spheres
with density $n\sigma^3=0.5$ and reduced shear rate $at_00=0.5$ where $t_00$ is the Boltzmann 
collision time. The circles are from simulation and the line is the GMSA result.}
\label{fig2}
\end{figure}

\begin{figure}[t]
\begin{center}
\leavevmode
\epsfxsize=5in 
\epsfbox{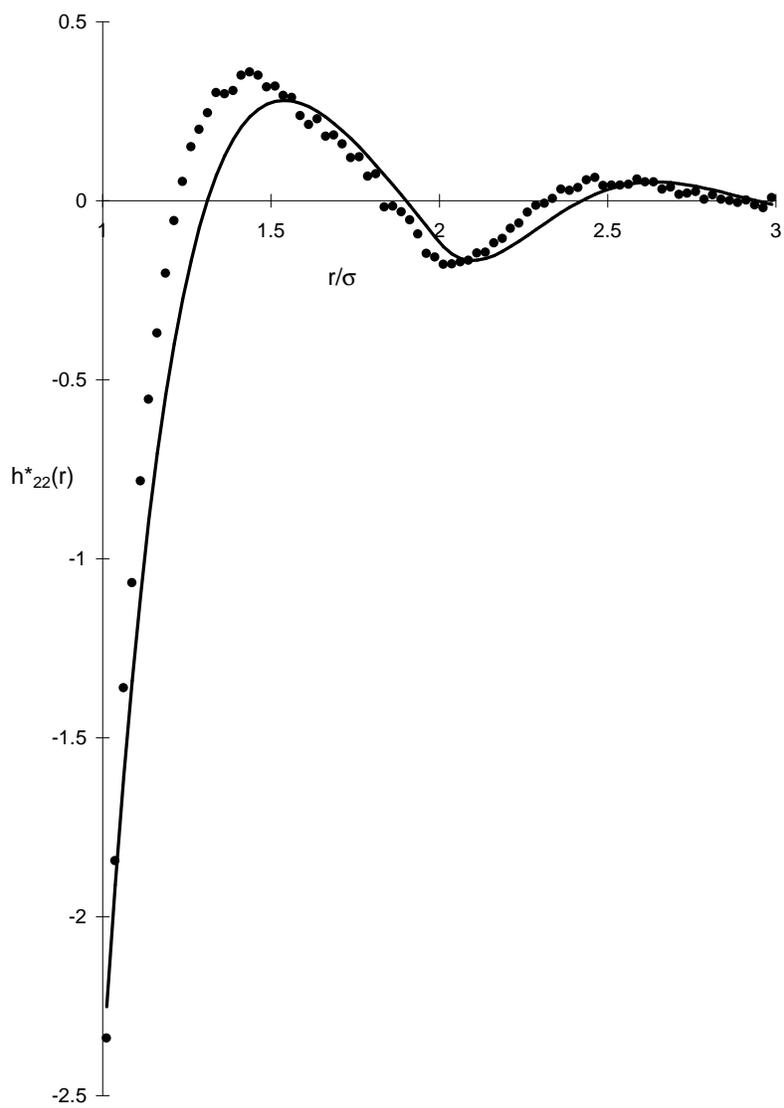}
\end{center}
\caption{Same as Fig. 2 except showing the quantity $h^{*}_{22}=i\sqrt{\frac{15}{2\pi }}h_{22}(r)$.}
\label{fig3}
\end{figure}
\end{document}